\documentclass[epj,nopacs]{svjour}

\usepackage{epsfig}
\usepackage{graphicx}
\usepackage{amsmath}
\usepackage{amssymb}
\usepackage{bm}

\newcommand{\be}{\begin{equation}}
\newcommand{\ee}{\end{equation}}
\newcommand{\bea}{\begin{eqnarray}}
\newcommand{\eea}{\end{eqnarray}}
\newcommand{\nn}{\nonumber}
\newcommand{\dst}{\displaystyle}
\newcommand{\fr}[2]{\frac{{\dst #1}}{{\dst #2}}}
\newcommand{\dd}{{\rm d}}

 \journalname{EPJ C}
 
\begin{document}

\setcounter{page}{1}
\title{
Large contribution of virtual Delbr\"uck scattering\\
to the emission of photons by relativistic nuclei\\ in
nucleus-nucleus and electron-nucleus collisions}
\titlerunning{Large contribution of the virtual Delbr\"uck scattering to
emission of photons }
\authorrunning{I.F.~Ginzburg et al.}
\author{I.F.~Ginzburg \inst{1,}\thanks{e-mail: ginzburg@math.nsc.ru} %
\and U.D.~Jentschura \inst{2,3,}\thanks{e-mail: Ulrich.Jentschura@mpi-nd.mpg.de} %
\and V.G.~Serbo \inst{4,2,}\thanks{e-mail:
serbo@math.nsc.ru}} \institute{{\addrNOVOSIBISKSO} \and
{\addrHDMPI} \and 
{\addrHDUNI} \and {\addrNOVOSIBIRSKST}}

\date{Received: September 6, 2007}


\newcommand{\addrHDUNI}{Institut f\"ur Theoretische Physik,
Philosophenweg 16, 69120 Heidelberg, Germany}

\newcommand{\addrNOVOSIBISKSO}{Sobolev Institute of Mathematics, 630090
Novosibirsk, Russia}

\newcommand{\addrNOVOSIBIRSKST}{Novosibirsk State University, 630090
Novosibirsk, Russia}

\newcommand{\addrHDMPI}{Max--Planck--Institut f\"ur Kernphysik,
Postfach 103980, 69029 Heidelberg, Germany}

\abstract{ Delbr\"uck scattering is an elastic scattering
of a photon in the Coulomb field of a nucleus via a virtual
electron loop. The contribution of this virtual subprocess
to the emission of a photon in the collision of
ultra-relativistic nuclei $Z_1\,Z_2\to Z_1\,Z_2\,\gamma$ is
considered. We identify the incoming virtual photon as
being generated by one of the relativistic nuclei involved
in the binary collision and the scattered photon as being
emitted in the process. The energy and angular
distributions of the photons are calculated. The discussed
process has no infrared divergence. The total cross section
obtained is 14 barn for Au--Au collisions at the RHIC
collider and 50 barn for Pb--Pb collisions at the LHC
collider. These cross sections are considerably larger than
those for ordinary tree-level nuclear bremsstrahlung in the
considered photon energy range $m_e \ll E_\gamma \ll m_e \,
\gamma$, where $\gamma$ is the Lorentz factor of the
nucleus. Finally, photon emission in electron-nucleus
collisions $e\,Z\to e\,Z\,\gamma$ is discussed in the
context of the eRHIC option.}


\maketitle

%
%
\section{Introduction and main results}

Recently, electromagnetic processes in
ultra-relativistic nuclear collisions were discussed in
numerous papers (see the review~\cite{BHTSKh-02} and references
therein) which is connected mainly with the operation of the
RHIC collider and the future LHC lead-lead option. For
these colliders the charge numbers of nuclei $Z_1=Z_2\equiv
Z$ and their Lorentz factors $\gamma_1=\gamma_2\equiv
\gamma$ are given in Table ~\ref{t1},
which is cited here from Ref.~\cite{PPP-06}.

\begin{table}[ht]
\begin{minipage}{8.0cm}
\vspace{5mm} {\renewcommand{\arraystretch}{1.5}
\caption{\label{table1} Colliders and cross sections for the
$Z\, Z\to Z\,Z\,\gamma$ process via the Delbr\"uck scattering
subprocess.}
\begin{center}
\par
\begin{tabular}{c@{\hspace{0.7cm}}c@{\hspace{0.4cm}}%
c@{\hspace{0.4cm}}c}
\hline
\hline
Collider & $Z$ & $\gamma$ & $\sigma^{}_{\rm }$ [barn] \\
\hline RHIC, Au--Au & 79 & 108 & 14
\\ \hline
LHC, Pb--Pb & 82 & 3000 & 50
\\
\hline
\hline
\end{tabular}
\label{t1}
\end{center}
}
\end{minipage}
\end{table}

Strictly speaking, only a few electromagnetic processes
with the production of leptons or photons are related to
fundamental physics. Nevertheless, many of them are of
imminent importance for two reasons: they are either
``dangerous,'' e.\,g.~in terms of possible beam losses, or
they are by contrast quite useful for experiments at the
RHIC and LHC colliders. These statements may be illustrated
by two examples.

{\it (i) $e^+e^-$ pair production} $Z_1Z_2\to
Z_1Z_2e^+e^-$. In a typical ultra-relativistic
collision, the number of the produced electrons is so
huge that some of them can be captured by nuclei, which
immediately leads to loss of these nuclei from the beam.
This capture process is an essential limitation for
the life time of the beam and determines
the maximal luminosity of a machine (for
details see Ref.~\cite{BHTSKh-02}).

{\it (ii) Coherent bremsstrahlung}. For the usual
brems\-strah\-lung the number of photons emitted in a single
collision of bunches is proportional to the number of particles in
the first and second bunches: $\dd N_{\gamma} \propto N_1 N_2\,
\dd E_{\gamma}/E_{\gamma}$. But when the photon energy decreases,
the coherence length becomes comparable to the length of the
second bunch, and radiation is caused by the interaction of a
nucleus $Z_1$ with the second bunch as a whole, but not with each
nucleus $Z_2$ separately, i.\,e.~coherent bremsstrahlung is an
emission of photons by particles of one bunch in the collective
electromagnetic field of the oncoming bunch. In this case $\dd
N_\gamma$ becomes proportional to the number of particles in the
first bunch and to the {\em squared} number of particles in the
second bunch: $\dd N_{\gamma} \propto N_1 N^2_2\, \dd E_{\gamma}/
E_{\gamma}$. As a result, the number of the produced photons at
RHIC becomes so huge in the infrared region that this process can
be used for monitoring beam collisions (see Ref.~\cite{ESS-96}).

In general, these considerations imply that various
electromagnetic processes have to be estimated (their cross
sections and distributions) in order not to miss some
interesting or potentially dangerous effects.

\begin{figure}[htb]
\includegraphics[width=1.0\linewidth]{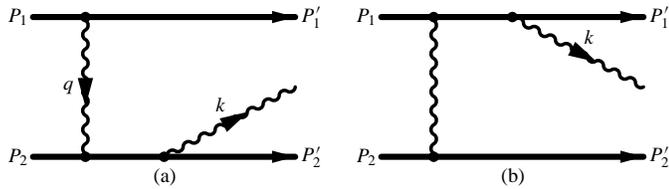}
\caption{Ordinary nuclear bremsstrahlung is the
emission of a photon in a nuclear collision via a
virtual Compton subprocess.}
 \label{Fig_1}
\end{figure}

Let us consider an emission of photons in an elastic nuclear
collision (i.\,e., without excitation of the nuclei in their final
state). The ordinary nuclear bremsstrahlung has been known in
detail for many years (see, for example, the review~\cite{BB-88}).
It is described by the Feynman diagrams of Fig.~\ref{Fig_1}{\sl a}
and Fig.~\ref{Fig_1}{\sl b}, in which the virtual photon is
emitted by either one of the nuclei and then this photon is
Compton scattered by the oncoming nucleus [In this as well as in
other figures we display not all but representative diagrams only.
E.\,g., in~Fig.~\ref{Fig_1}{\sl a}, the diagram with the exchange
of photons with momenta $q$ and $k$ have to be added to complete
the entire range of diagrams.]

%
%

\begin{figure}[htb]
\includegraphics[width=1.0\linewidth]{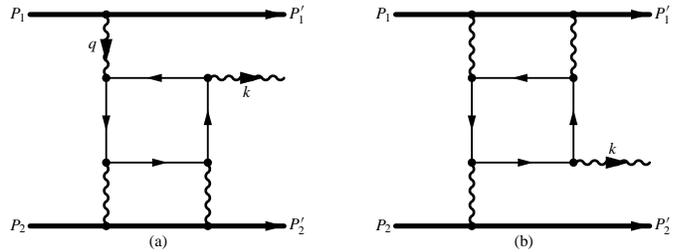}
\caption{Emission of a photon in a nuclear collision via
virtual Delbr\"uck scattering in the lowest order of
quantum electrodynamics (QED). Bold lines denote nuclei,
thin lines denotes the electron propagators.}
 \label{Fig_2}
\end{figure}

In the present paper, we consider in detail the emission of
photons not due to the Compton subprocess, but due to
another one -- namely, the Delbr\"uck scattering subprocess
of Fig.~\ref{Fig_2}. Delbr\"uck scattering is an elastic
scattering of a photon in the Coulomb field of a nucleus
via a virtual electron loop. To the best of our knowledge,
the process of Fig.~\ref{Fig_2} has been mentioned for the
first time in the paper~\cite{BB-ZfP-88}, but its cross
section has not yet been calculated or estimated. However,
the situation is interesting because for heavy nuclei, the
cross section of Delbr\"uck scattering is known to be
greater by one order of magnitude than for Compton
scattering of photons with small energies. Therefore, we
expect that the photon emission of Fig.~\ref{Fig_2} due to
Delbr\"uck scattering should be dominant over the ordinary
bremsstrahlung of Fig.~\ref{Fig_1} in a certain range of
photon energies.

%
%

At first sight, the process of Fig.~\ref{Fig_2} looks like
a typical quantum electrodynamic (QED) loop correction to
the Compton scattering and, therefore, should have a small
cross section $\sigma \propto \alpha^7$, where
$\alpha\approx 1/137$ is the fine structure constant.
However, at second sight, we should add a very large factor
$Z^6\sim 10^{11}$ (we assume a collision of identical
nuclei with $Z = Z_1 = Z_2$) and take into account that the
natural scale of the cross section is the square of the
electron Compton wavelength $\hbar^2/(m_e\,c)^2$, where
$m_e$ is the electron mass, $\hbar$ is the Planck constant
and $c$ is the light velocity. And last, but not least, we
show below that this cross section has an additional
logarithmic enhancement of the order of
 \be
L^2 \gtrsim 100\,, \;\;\;
L=\ln{\left(\gamma_1\gamma_2\right)}\,.
 \label{1}
 \ee

As a result, the discussed cross section is of the order of
 \be
\sigma \sim \alpha \, \left(Z\alpha\right)^6 \,
\,\fr{\hbar^2}{m_e^2\, c^2}\,L^2 \,,
 \ee
which gives $130 \, {\rm barn}$ for Pb--Pb at LHC. A more
detailed calculation, as reported below, gives a result
which differs from that estimate only by a numerical
prefactor approximately equal to $0.4$. The corresponding
cross sections are given in Table~\ref{table1}. In
particular, for the LHC collider,
 \be
\sigma =  50 \, {\rm barn}\,.
 \ee
Note for comparison, that this cross section is 6 times
larger than for the total hadronic/nuclear cross section in
Pb--Pb collisions, which is roughly 8~barn.

\begin{figure}[htb]
\includegraphics[width=1.0\linewidth]{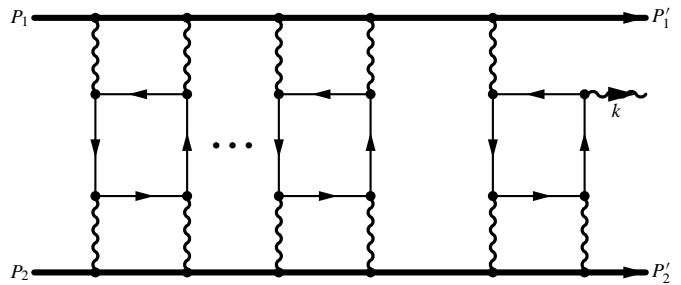}
\caption{Feynman diagram for the unitarity correction.}
 \label{Fig_3}
\end{figure}

To complete the description, we should mention that there
is a numerically not large, but conceptually interesting
so-called unitarity correction to the discussed process. It
is due to the unitarity requirement for the $S$ matrix and
corresponds to the exchange of light-by-light blocks
between nuclei (Fig.~\ref{Fig_3}); this correction is
analyzed in detail in~\cite{in_preparation}, where it is
shown that the unitarity correction can be estimated by the
simple expression $\sim -0.5\,(Z\alpha)^4$.

The obtained results can be easily generalized to the
photon emission in electron-nucleus collisions without
excitation of nucleus $e\, Z \to e \, Z \, \gamma$. This
consideration is motivated by project of the eRHIC collider
which is now actively discussed as a promising extension of
the existing RHIC machine (see~\cite{eRHIC}). In
particular, the proposal is to build an additional electron
ring with the energy $E_e= 10$~GeV and, thus, to create an
electron-nucleus collider. Certainly, the emission of
photons in the direction of the electron beam is dominated
by ordinary bremsstrahlung. By contrast, for the emission
in the {\em nuclear} beam direction, we find out that the
process via Delbr\"uck scattering is dominant in a certain
region of the photon energy.

The paper is organized as follows. In Sec.~\ref{delbrueck},
we call some known properties of Delbr\"uck scattering. In
Sec.~\ref{scatteringZZ}, we calculate the contribution of
Delbr\"uck scattering to the cross section of the $Z_1 \,
Z_2 \to Z_1 \, Z_2 \, \gamma$ process.
Sec.~\ref{comparisons} is devoted to a quantitative
comparison of our process to other, competing photon
emission processes. In Sec.~\ref{scatteringeZ} we discuss
the case of the electron-nucleus collisions. A final
conclusions are given in Sec.~\ref{conclusions}. A short
summary of our main results is presented in
Ref.~\cite{GJS-Letters}.

Throughout this text, we use natural units with $\hbar = c
= 1$ and $\alpha\approx 1/137$, and we denote the electron
(nucleus) mass by $m\equiv m_e$ ($M$), respectively.
The scalar product of two 4-vectors is
$P_1\cdot P_2=E_1E_2-{\bm P}_1{\bm P}_2$.

\begin{figure}[htb]
\includegraphics[width=0.7\linewidth]{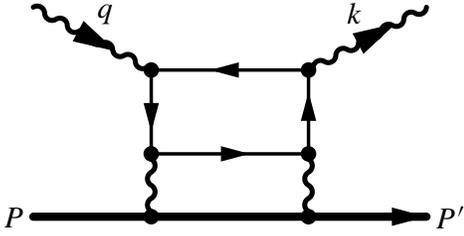}
\caption{The block Feynman diagram for Delbr\"uck
scattering in the lowest QED order.}
 \label{Fig_4}
\end{figure}

%
%
\section{Basics of Delbr\"uck scattering}
\label{delbrueck}

Delbr\"uck scattering is one of a few fundamental
non-linear QED processes which had been studied
experimentally. It can be described as a transition, in
which the initial real photon fluctuates into a virtual
electron-positron pair, interacts with a nucleus and then
transforms back to the final photon, i.\,e.~it is an
elastic scattering of a photon in the Coulomb field of a
nucleus via a virtual electron loop. In one-loop order, it
is described by the block diagram of Fig.~\ref{Fig_4}.
Properties of Delbr\"uck scattering are well known (see,
e.\,g., the review~\cite{Milstein-Schumachar-94}, recent
experiments~\cite{BINP} and numerical results for the
Delbr\"uck scattering amplitudes in Ref.~\cite{F-92}). The
total cross section of this process $\sigma_{\rm
D}(\omega_L, Z)$ depends on the invariant (see
Figs.~\ref{Fig_2} for the identification of $q$ and $P$)
 \be
\omega_L=\fr{q\cdot P}{M}\,,
  \label{4}
  \ee
which is equal to the initial photon energy in the
laboratory system (lab-system, denoted by the
subscript~$L$). Here the lab-system means the rest frame of
the scattering nucleus, in which the 4-momentum of the
initial photon takes the form $q=(\omega_L,0,0,\omega_L)$,
and the 4-momentum of the initial nucleus is $P=(M,0,0,0)$.
This cross section vanishes at small energies,
 \be
\sigma_{\rm D}(\omega_L \ll m, Z)\sim (Z\alpha)^4\,
{\alpha^2\over m^2}\, \left(\fr{\omega_L}{m}\right)^4 \,,
 \label{5}
 \ee
and tends to a constant, independent of $\omega_L$, in the
limit $\omega_L\gg m$. In the lowest order of the QED
perturbation theory, this constant is
 \be
\sigma_{\rm D}(\omega_L \gg m, Z) = \sigma_{\rm D}^{(0)}(Z)
=1.07\, (Z\alpha)^4\, {\alpha^2\over m^2} \,.
 \label{6}
 \ee
For heavy nuclei, the strong-field effects (so-called
Coulomb corrections) drastically change this result. These
corrections correspond to the exchange of virtual photons
between the electron loop and the nucleus. They are of even
order in powers of $Z\alpha$ (due to the Furry theorem for
the electron loop), i.\,e. they are proportional to
$(Z\alpha)^{2n}$ and decrease significantly this constant,
 \be
\sigma_{\rm D}^{(0)}(Z) \to\sigma_{\rm D}(Z) \equiv
{\sigma_{\rm D}^{(0)} (Z)\over r_Z}\,,
 \label{7}
 \ee
where the reduction factor $r_Z > 1$. For example, for the
Delbr\"uck scattering off the Au ($Z=79$) and Pb ($Z=82$)
nuclei, it is $r_{79}=1.7$ and $r_{82}=1.8$, respectively,
and the corresponding cross sections read as follows
\begin{subequations}
\begin{align}
\sigma_{\rm D}(Z=79) =& \; 5.5 \times 10^{-3}
 \;{\rm barn}\,,\;\;
\\[2ex]
\sigma_{\rm D}(Z=82) =& \; 6.2 \times 10^{-3} \;{\rm barn}
\,.
\end{align}
\label{8}
\end{subequations}

It should be noted that the cross section for the Delbr\"uck
scattering off heavy nuclei is considerable larger than that
for the nuclear Thomson scattering (which is the low-energy
limit of Compton scattering)
 \be
\sigma_{\rm T}(Z) = {8\pi\over 3} {Z^4 \alpha^2\over M^2}\,,
 \label{9}
 \ee
where $M$ is the mass of nucleus. Indeed, the ratio
 \be
{\sigma_{\rm T}(Z)\over \sigma_{\rm D}(Z)}= 7.83\, r_Z \,
\left({m\over \alpha^2 \, M}\right)^2 \approx {1\over 30}
\;\;{\rm for\;Au\;and\;Pb}\,.
 \label{10}
 \ee

For the analysis to be described below,
we will also need the differential cross section
of Delbr\"uck scattering over the transverse momentum
of the final photon $k_\perp$ in the region $\omega_L \gg
m$. The nucleus can be considered as a point-like particle
up to the limit imposed by the nuclear form factor, i.\,e. up
to $k_\perp \sim 1/R$, where $R \approx 1.2\,A^{1/3}$~fm is
the radius of the nucleus with $A$ the nucleon number (for both
Au and Pb nuclei, we have $R \approx 7$~fm and $1/R \approx
28$~MeV). In this region, the differential cross section
for Delbr\"uck scattering can be written in the form
 \be
\dd \sigma_{\rm D} = \alpha^2\, (Z\alpha)^4\,
f_Z(k_\perp/m) \;{\dd k_\perp^2\over m^4_\perp}
 \label{11}
 \ee
with
 \be
m_\perp = \sqrt{m^2+k_\perp^2}\,.
 \ee
The function $f_Z(k_\perp/m)$ is of the order of unity at
$k_\perp/m \lesssim 1$ (except for the region of very small
$k_\perp/m\lesssim m/\omega_L$), and it is a slowly varying
function at larger values of $k_\perp/m$. Numerical values
of this function can be found from plots and numbers given
in Refs.~\cite{Milstein-Schumachar-94,F-92}, in particular,
 \be
f_{82}(k_\perp/m=1) \approx 0.48\,,\;\;f_{82}(k_\perp/m\gg
1)\approx 1.2\,.
 \ee
As a result, we can conclude that the main contribution to
the total cross section at  $\omega_L \gg m$ [given in
Eqs.~(\ref{6})--- (\ref{8})] comes from the region where
the transverse momenta of the final photon are of the order
of the electron mass, $k_\perp \sim m$.

%
%
\section{Delbr\"uck scattering and
the $\boldsymbol{ Z_1\, Z_2\to Z_1\, Z_2 \, \gamma }$
process}
 \label{scatteringZZ}

Let us consider, in general terms, the process of a photon
emission without any excitation of nuclei in the final
state:
 \be
Z_1(P_1)+Z_2(P_2) \to Z_1(P^\prime_1)+Z_2(P^\prime_2)
+\gamma(k)\,.
 \label{12}
 \ee
In this process two nuclei with charges $Z_1e$ and $Z_2e$
and 4-momenta $P_1$ and $P_2$ collide with each other and
produce a photon with the total four--momentum $k$. Let
$E_i$ ($\gamma_i=E_i/M_i$) and $E_\gamma$ be the energy
(the Lorentz factor) of the $i$th nucleus and the photon
energy in the collider system, respectively. Here, the
collider system means the rest frame of the particle
collider, which is not necessarily equal to our
``lab-system'' (the latter we define to be the rest frame
of the scattering nucleus). The collider system, by
contrast, coincides with the center-of-mass system for
identical colliding nuclei, and in this case $\gamma_1 =
\gamma_2=\gamma$.

\begin{figure}[htb]
\includegraphics[width=0.8\linewidth]{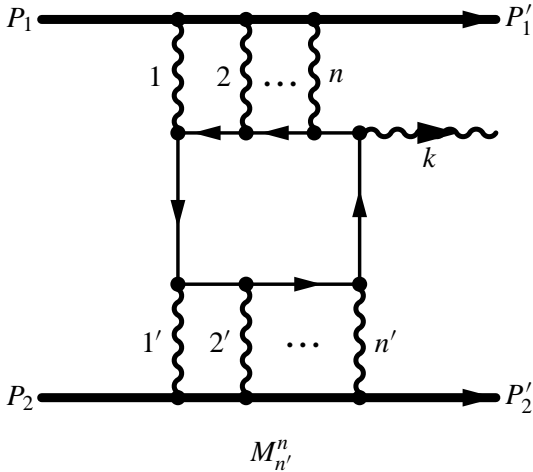}
\caption{Amplitude  $M^n_{n'}$ for the emission of a photon
in the nuclear collision; here $n (n')$ is the number of
exchange virtual photons between the electron loop and the
first (second) nucleus.}
 \label{Fig_5}
\end{figure}

%
%
\subsection{Total cross section}

The contribution of the Delbr\"uck scattering subprocess to
the cross section of the process (\ref{12}) is described in
the lowest QED order by the Feynman diagrams displayed in
Fig.~\ref{Fig_2}. However, the parameter of the
perturbation series $Z\alpha$ is of the order of unity for
heavy nuclei. For example, $Z\alpha \approx 0.6$ for the
discussed colliders. This means that other amplitudes
$M^n_{n'}$ (see Fig.~\ref{Fig_5}) with the exchange of $n
(n')$ virtual photons between electron loop and the first
(second) nucleus have to be taken into account. Therefore,
the whole series in $Z\alpha$ has to be summed to obtain
the cross section with sufficient accuracy. Fortunately,
there is another small parameter
$\eta \equiv 1/\ln(\gamma_1\gamma_2)$, and it will be sufficient to
calculate the cross section in the leading logarithmic
approximation (LLA), where the omitted terms are of
the order of
 \be
\eta = \fr{1}{\ln(\gamma_1\gamma_2)} =0.1 \;\mbox{for
RHIC and 0.06 for LHC}\,.
 \ee

\begin{figure}[htb]
\includegraphics[width=1.0\linewidth]{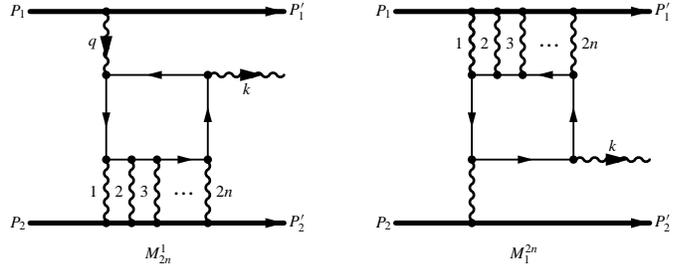}
\caption{Amplitudes  $M^1_{2n}$ ($M^{2n}_1$) with a single
exchange photon between the electron loop and the first
(second) nucleus.}
 \label{Fig_6}
\end{figure}

Let ${\cal M}$ be the sum of the amplitudes $M^n_{n'}$ of
Fig.~\ref{Fig_5}. This sum can be presented in the form
\begin{eqnarray}
\label{7a} {\cal M}&=& \sum_{n\,n'\geq 1} M^n_{n'}=
M_1+\tilde{M}_1+ M_2\,,\\
M_1 &=& \sum_{n\geq 1} M^1_{2n}\,, \ \ \tilde{M}_1 =
\sum_{n\geq 1} M^{2n}_1\,, \ \ M_2= \sum_{n\, n'\geq 2}
M^n_{n'} \,. \nonumber
\end{eqnarray}
The amplitude $M_1$ contains a one-photon exchange with
the first nucleus and multiple photon exchanges with the
second nucleus, whereas the amplitude $\tilde{M}_1$
describes a one-photon exchange with the second nucleus
and multiple photon exchanges with the first nucleus
(Fig.~\ref{Fig_6}). In the last amplitude $M_2$, there is
no one-photon exchange, it describes a multi-photon
exchange between electron loop and both nuclei. According
to this classification, we write the total cross section as
\begin{equation}
\sigma = \sigma_1 +\tilde\sigma_1 + \sigma_2\,,
 \label{14}
\end{equation}
where
\begin{eqnarray}
\dd \sigma_1&\propto& |M_1|^2 \,,\;\;\; \dd \tilde\sigma_1
\propto |\tilde{M}_1|^2 \,,\;
\\
\dd \sigma_2& \propto& 2 \, {\rm Re}\left( M_1 \tilde{M}_1^* +M_1
M_2^* +\tilde{M}_1 M_2^* \right) + |M_2|^2 \,.
 \nonumber
\end{eqnarray}

The integration over the transferred momentum squared $q^2$
results in the large Weiz\-s\"acker--Williams logarithm
$\sim L$ for $\sigma_1$, with the same being true for
$\tilde\sigma_1$. The contribution $\sigma_2$ does not contain such
a logarithm. Therefore, the relative contribution of the
$\sigma_2$ term is
\be
\fr{\sigma_2}{\sigma_1} \sim \fr{(Z\alpha)^2}{L}< 0.04\,.
\ee
As a result, with an accuracy of the order of a few percent
we can neglect $\sigma_2$ in the total cross section and
use the equation
\begin{equation}
\sigma =  \sigma_1 +\tilde\sigma_1\,.
 \label{17}
\end{equation}

Let us consider the cross section $\sigma_1$, within the
four-momentum conventions of Fig.~\ref{Fig_6}. In the LLA,
it can be calculated using the equivalent photon
approximation, in which $\dd \sigma_1$ is expressed via the
number of equivalent photons $\dd n_1$, emitted by the
first nuclei, and the cross section for the Delbr\"uck
scattering off the second nuclei (see, e.g.,
Ref.~\cite{BGMS}):
 \begin{equation}
\dd \sigma_1 = \dd n_1\,\, \sigma_{\rm D}(\omega_L, Z_2) \,.
 \label{18}
\end{equation}
The virtual Delbr\"uck scattering amplitude decreases when the
virtuality of the initial photon $Q^2 =-q^2$ becomes larger
than $m_\perp^2$ (here, $q=P_1-P_1^\prime$ is the 4-momentum
of the equivalent photon). This means that the main
contribution to $\dd \sigma_1$ is given by photons from the
first nucleus with a small virtuality
 \be
Q^2 =-q^2={\bf q}_\perp^2
+\left(\fr{\omega}{\gamma_1}\right)^2 \ll m_\perp^2\,,
 \label{19}
 \ee
where $\omega=E_1-E_1^\prime$ is the energy of the
equivalent photon in the collider system. Therefore, we can
neglect the virtuality of this photon in the description of
the cross section $\sigma_{\rm D}(\omega_L, Z_2)$ for the
subprocess. From~(\ref{19}), we learn that we can usually
assume $\omega \ll m_\perp \gamma_1$. Because
$\omega_L$ can be expressed in terms of
Lorentz invariants as
 $$
\omega_L=(q\cdot P_2)/M_2=2\omega \gamma_2\,,
 $$
the most important region for this cross section is [in
accordance with Eqs. (\ref{5})---(\ref{8}) and (\ref{19})]
\be
m \ll \omega_L=2 \, \omega \, \gamma_2 \ll m \, \gamma_1 \, \gamma_2\,,
\qquad
k_\perp \sim m\,.
\label{20}
\ee

To calculate the spectrum of equivalent photons, we can use
Eq.~(D.4) from Ref.~\cite{BGMS} neglecting terms
proportional to $\omega /E_1$, since in our case
$\omega \lesssim m \gamma_1 \ll E_1$:
 \begin{equation}
\dd n_1(\omega, Q^2) = {Z_1^2 \alpha \over \pi}\, {\dd \omega \over
\omega}\, \left( 1- {Q^2_{\min} \over Q^2 } \right) \,F^2(Q^2) \,
{\dd Q^2\over Q^2} \,,
 \label{21}
 \end{equation}
where
\begin{equation}
Q^2_{\min} = {\omega^2\over \gamma_1^2}\,,
\end{equation}
and $F(Q^2)$ is the nuclear electromagnetic form factor.
The function $F(Q^2)$ is normalized by the ``charge conservation''
condition $F(0)=1$,
but drops very quickly at $Q^2$ larger than $1/R^2$. In
this section, we assume that $m_\perp^2$ is considerable
smaller than $1/R^2$. This implies, in particular,
that $k_\perp \ll 1/R$, and the nucleus is effectively
probed in the long-wavelength limit,
so that we can put $F(Q^2)=1$ in our
calculation. Integrating this number over $Q^2$ in the
region
 \be
Q^2_{\min}= \frac{\omega^2}{\gamma_1^2}
\leq Q^2 \lesssim m^2\,,
 \label{22}
 \ee
we obtain the spectrum of equivalent photons as
 \be
\dd n_1(\omega) = 2\,{Z_1^2 \alpha \over \pi}\, {\dd\omega
\over \omega}\, \ln\left( \fr{m\gamma_1}{\omega} \right)\,.
 \label{23}
 \ee

Then integrating the cross section (\ref{18}) over $\omega$
in the region
 \be
\fr{m}{\gamma_2}\lesssim \omega \lesssim m \gamma_1\,,
 \label{24}
 \ee
we obtain the total cross section $\sigma_1$ in the LLA
\begin{align}
\sigma_1 =& \; \fr{Z_1^2 \alpha}{\pi}\,
\sigma_{\rm D}(Z_2)\,L^2\,,\;\;
\nonumber\\[2ex]
L =& \; \ln\left( \fr{P_1 \cdot P_2}{2M_1M_2} \right) =
\ln(\gamma_1 \gamma_2)\,.
\label{25}
\end{align}
Analogously, the cross section $\tilde\sigma_1$ is
 \be
\tilde\sigma_1= \fr{Z_2^2 \alpha}{\pi}\, \sigma_{\rm
D}(Z_1)\,L^2\,.
 \label{26}
 \ee

As a result, the total contribution of virtual Delbr\"uck
scattering to the cross section of the process given by
formula~(\ref{12}) is equal to
\begin{align}
\sigma =& \; \sigma_1 + \tilde\sigma_1=
\fr{\alpha}{\pi}\,
\left[Z_1^2 \, \sigma_{\rm D}(Z_2)\,+
      Z_2^2 \, \sigma_{\rm D}(Z_1)\right] \, L^2\,,
\label{27}
\end{align}
where $L$ is given in Eq.~(\ref{25}) and $\sigma_{\rm
D}(Z)$ in Eqs.~(\ref{7})---(\ref{8}). In particular, for
Au--Au collisions at the RHIC collider, the total cross
section is $\sigma= 14$~barn, and for the Pb--Pb collisions
at the LHC collider, the total cross section $\sigma=
50$~barn (see also~Table~\ref{table1}).

%
%
\subsection{Energy and angular distribution of final photons}

In the previous subsection, we have considered only the
total cross section. Here, we are interested in the angular
distribution of the final photons, and therefore, the
dependence on $k_\perp$ has to be restored. Let us
therefore consider the region of not too small transverse
momenta, \be m\lesssim k_\perp \lesssim  1/R\,, \label{32}
\ee which differs from the condition $k_\perp \ll 1/R$
employed in the previous calculation. To obtain the number
of equivalent photons in this region, we should integrate
expression (\ref{21}) not in the interval (\ref{22}), but
in the larger interval
 \be
Q^2_{\min}=(\omega/\gamma_1)^2\leq Q^2 \lesssim m_\perp^2\,,
 \label{33}
 \ee
whose upper limit is $m_\perp^2$, not $m^2$.
This leads to
 \be
\dd n_1(\omega) = 2\,{Z_1^2 \alpha \over \pi}\, {\dd\omega
\over \omega}\, \ln{\fr{m_\perp\gamma_1}{\omega}}\,.
 \label{34}
 \ee
Using this expression and the distribution of the
Delbr\"uck subprocess [given by Eq. (\ref{11})], we obtain
the differential cross section
 \be
\dd \sigma_1= \fr{2}{\pi}\alpha
(Z_1\alpha)^2(Z_2\alpha)^4\,
\ln\left({\fr{m_\perp\gamma_1}{\omega}}\right)\,
f_{Z_2}(k_\perp/m){\dd\omega \over \omega}\,{\dd
k_\perp^2\over m_\perp^4}\,,
 \label{35}
 \ee
which is valid in the region
 \be
\fr{m_\perp}{\gamma_2} \ll  \omega \ll m_\perp \, \gamma_1
\,, \qquad k_\perp \sim m_\perp\,.
 \label{20a} \ee
To find the energy and angular distribution of the final
photons, we should now express the energy of the equivalent
photon $\omega$ in this formula in terms of the energy of
the final photon $E_\gamma$.

In the main region for the cross section $\sigma_1$, the
energy of the final photon in the rest frame of the second
nucleus
 $$
E_{\gamma L}= \fr{(k \cdot P_2)} {M_2}
=\omega_L-\fr{k^2_\perp}{2M_2}
 $$
and the longitudinal momentum (in the $z$ direction, as measured
in the lab frame $L$)
 $$
k_{zL} = E_{\gamma L} - \fr{k_\perp^2}{2E_{\gamma L}}
 $$
almost coincide with $\omega_L$. Going to the collider
frame of reference, we find the energy of the final photon
\be E_\gamma = \gamma_2 \, \left(E_{\gamma L} -
V_2\,k_{zL}\right) = \omega+\fr{k_\perp^2}{4 \, \omega}
\label{36} \ee and its longitudinal momentum \be k_z =
\gamma_2 \, \left(k_{zL}- V_2\,\omega_L\right) =
\omega-\fr{k_\perp^2}{4 \, \omega}\,, \label{37} \ee where
$\gamma_2= 1/\sqrt{1-V_2^2}$ and $V_2$ is the velocity of
the second nucleus in the collider system.

For further analysis, it will be useful to split the
region~(\ref{20a}) into two subregions with small and large
energies of the equivalent photon:
 \be
{\rm subregion}\; A:\;\;\; \fr{m_\perp}{\gamma_2}\ll \omega
\ll m_\perp\,, \label{38}
 \ee
and
 \be
{\rm subregion}\; B:\;\;\; m_\perp\ll \omega \ll m_\perp
\gamma_1\,.
 \label{39}
 \ee

Now we can see that the longitudinal momentum of the final
photon is positive in the region $B$,
 \be
k_z\approx E_\gamma \approx \omega\,.
 \label{40}
 \ee
Therefore, the photon flies along the momentum of the first
nucleus $\bm{P}_1$, and
\begin{align}
& \dd \sigma^{(B)}_1  = \fr{2}{\pi^2} \alpha (Z_1\alpha)^2
(Z_2\alpha)^4 \fr{f_{Z_2}(k_\perp/m)}
{\left(m^2+k_\perp^2\right)^2} \ln\left(
{\fr{m_\perp\gamma_1}{E_\gamma}} \right)\,
\fr{\dd^3k}{E_\gamma}\,,\nonumber\\[2ex]
& m_\perp\ll
E_\gamma\ll m_\perp\gamma_1 \,.
\label{41}
\end{align}
The contributions of this region into the spectrum of the
final photon is
\begin{align}
& \dd \sigma^{(B)}_1= \fr{2}{\pi}\,Z_1^2\alpha \,
\sigma_{\rm D}(Z_2)\,
\ln\left( {\fr{m\gamma_1}{E_\gamma}} \right)\;
\fr{\dd E_\gamma}{E_\gamma}\,,
\nonumber\\[2ex]
& m\ll E_\gamma\ll m \, \gamma_1 \,.
\label{42}
\end{align}

On the contrary, the longitudinal momentum of
the final photon is negative in subregion $A$:
 \be
k_z\approx -E_\gamma \approx -\fr{k_\perp^2}{4\omega}\,.
 \label{43}
 \ee
Therefore, the photon flies along the momentum of the second
nucleus $\bm{P}_2=-\bm{P}_1$ and
\begin{align}
& \dd \sigma^{(A)}_1= \fr{2}{\pi^2}\, \alpha (Z_1\alpha)^2
(Z_2\alpha)^4 \fr{f_{Z_2}(k_\perp/m)}
{\left(m^2+k_\perp^2\right)^2} \ln\left( \fr{\gamma_1 \,
E_\gamma}{m_\perp} \right)\; \fr{\dd^3k}{E_\gamma}\,,
\nonumber\\[2ex]
& m_\perp\ll E_\gamma\ll m_\perp\gamma_2 \,.
\label{44}
\end{align}
The contributions of this region to the spectrum of the
final photons is
\begin{align}
& \dd \sigma^{(A)}_1 =
\fr{2}{\pi}\,Z_1^2\alpha
\sigma_{\rm D}(Z_2)\,
\ln\left( {\fr{\gamma_1 E_\gamma}{m}} \right)\,
\fr{\dd E_\gamma}{E_\gamma}\,,
\nonumber\\[2ex]
& m\ll E_\gamma\ll m \, \gamma_2 \,.
\label{45}
\end{align}
One verifies that reassuringly, after integration over
$E_\gamma$, the sum of the two expressions (\ref{42}) and
(\ref{45}) coincides with $\sigma_1$ from (\ref{25}).

The corresponding expressions for $\dd\tilde\sigma_1$ can
be easily obtained. For the sum $\dd \sigma= \dd \sigma_1 +
\dd \tilde\sigma_1$, we present results only for the case
of identical nuclei and for the differential over the
photon momentum cross section
\begin{align}
& \dd \sigma = \; \fr{2}{\pi^2}\, \alpha (Z\alpha)^6
\fr{f_{Z}(k_\perp/m)} {\left(m^2+k_\perp^2\right)^2}\,
\,L\, \fr{\dd^3k}{E_\gamma}\,,
\nonumber\\[2ex]
& m_\perp\ll E_\gamma\ll m_\perp\gamma
 \label{46}
\end{align}
and for the spectrum of photons
 \be
\dd \sigma= \fr{4}{\pi}\, Z^2\alpha \, \sigma_{\rm
D}(Z)\,L\, \fr{\dd E_\gamma}{E_\gamma}\,, \qquad m\ll
E_\gamma\ll m\,\gamma \,.
 \label{47}
 \ee

The typical emission angle of the photon is not very small:
\be
\fr 1\gamma \ll \theta_\gamma=\fr{k_\perp}{E_\gamma}
\ll 1\,.
\ee
It is useful to make two observations:

{\it First remark}. The scattering angles of both nuclei in
the discussed process are very small but different for
specific contributions to $\sigma_1$ and $\tilde\sigma_1$.
In the case of~$\sigma_1$, the first nucleus gets a
transverse momentum $P_{1\perp}'= q_\perp\ll m$, which is
considerably smaller than a typical transverse momentum of
the second scattered nucleus $P_{2\perp}'\approx k_\perp
\sim m$. In the case of~$\sigma_2$, vice-a-vise, after the
scattering process, the first nucleus has a transverse
momentum which is considerable larger than that of the
second nucleus. That is the reason why the interference of
these two contributions is small.

{\it Second remark}. The energy distribution of
photons~(\ref{47}) has the form
 \be
\dd \sigma \propto \fr{\dd E_\gamma}{E_\gamma}\,,
 \ee
which is typical for the bremsstrahlung spectrum of soft
photons and usually leads to an infrared divergence for
the total cross section. In our case, this type of
distribution is only valid for not too soft photons in the
region $m\ll E_\gamma\ll m\gamma$. When the photon energy
tends to zero, we should take into account that the
Delbr\"uck cross section vanishes for soft photons [see Eq.
(\ref{5})]. As a result, the discussed cross section in
fact has no infrared divergence.

%
%
\section{Comparisons}
\label{comparisons}

%
%
\subsection{Comparison to ordinary nuclear bremsstrahlung}

The ordinary photon emission by nuclear bremsstrahlung is
described by the block Feynman diagrams of
Fig.~\ref{Fig_1}{\sl a} and Fig.~\ref{Fig_1}{\sl b}. Let
the cross section $\dd \sigma_{\rm br}^a$ and $\dd
\sigma_{\rm br}^b$ correspond to the diagrams of
Fig.~\ref{Fig_1}{\sl a} and Fig.~\ref{Fig_1}{\sl b},
respectively. The bremsstrahlung cross section is
\begin{equation}
\dd \sigma_{\rm br} = \dd \sigma_{\rm br}^a + \dd
\sigma_{\rm br}^b \,,
 \label{59}
\end{equation}
because the interference term is small and can be safely
neglected.

Now we can repeat the previous calculations with minor
changes. The expression analogous to (\ref{18}) has the
form
 \begin{equation}
\dd \sigma_{\rm br}^a = \dd n_1(\omega) \,
\dd \sigma_{\rm C}(\omega, Z_2) \,,
 \label{60}
\end{equation}
where $\dd n_1(\omega)$ is the number of the equivalent
photons emitted by the first nucleus, and $\sigma_{\rm C}$
is the cross section for the Compton scattering of this
photon off the second nucleus. In the calculation of $\dd
n_1(\omega)$, we should integrate the expression (\ref{21})
over $Q^2$ up to a limit, which is imposed by the decrease
of the form factor of the nucleus for large $Q^2 \gg
1/R^2$. Therefore,
 \be
\dd n_1(\omega) = 2\,{Z_1^2 \alpha \over \pi}\,
\ln\left( \fr{\gamma_1}{\omega R} \right)\,
{\dd \omega \over \omega}\,.
 \label{61}
 \ee
For the Compton cross section, we can use a well-known
expression valid for a charged point particle (such an
approach gives a good approximation at least in the region
of not too energetic photons),
 \bea
\dd \sigma_{\rm C}(\omega, Z_2)&=& \frac32\,
\sigma_{\rm T}(Z_2) \, \left(x-2x^2+2x^3\right)\,
\fr{\dd E_\gamma}{E_\gamma}\,,\nn\\
x&=& \fr{\omega_{\min}}{\omega}\,,\qquad
\omega_{\min}=\fr{E_\gamma}{4 \, \gamma_2^2}\,,
 \label{62}
 \eea
where the nuclear Thomson cross section $\sigma_{\rm T}(Z)$
is given in Eq.~(\ref{9}).

Integrating (\ref{60}) over $\omega$ from $\omega_{\min}$,
we find (the upper limit of this integration can be set
to infinity due to fast convergence of the integral)
 \be
\dd \sigma_{\rm br}^a=
\fr{2}{\pi}\,Z_1^2\alpha \,
\sigma_{\rm T}(Z_2)\,
\ln\left( \fr{4 \, \gamma_1 \gamma_2^2}{E_\gamma R} \right)\;
\fr{\dd E_\gamma}{E_\gamma}\,.
 \label{64}
 \ee
Analogously,
 \be
\dd \sigma_{\rm br}^b= \fr{2}{\pi}\,Z_2^2\alpha \,
\sigma_{\rm T}(Z_1)\, \ln\left( \fr{4\,\gamma_2 \,
\gamma_1^2}{E_\gamma \, R} \right)\; \fr{\dd
E_\gamma}{E_\gamma}\,.
 \label{65}
 \ee

To compare these cross sections with the corresponding ones
for the Delbr\"uck scattering, we consider the case of
identical nuclei and calculate the contribution of the
ordinary nuclear bremsstrahlung integrated over the region
$m< E_\gamma < m \gamma$, which is the main region for the
total cross section $\sigma$ from Eq. (\ref{27}). This
gives
\bea
\Delta\sigma_{\rm br} &=&
\fr{2}{\pi}\,Z^2 \, \alpha \sigma_{\rm T}(Z)\,
L \, \left(L+l_\gamma\right)\,,
\nonumber\\
l_\gamma &=& \ln\left( \fr{4\,\sqrt{\gamma}}{m R} \right).
\eea
As a result, we find that the ratio
 \be
\fr{\Delta\sigma_{\rm br}}{\sigma}= \fr{\sigma_{\rm
T}(Z)}{\sigma_{\rm D}(Z)}
\,\left(1+\fr{l_\gamma}{L}\right)
 \label{66}
  \ee
is indeed small:
 \be
\fr{\Delta\sigma_{\rm br}}{\sigma} \approx \fr{1}{15} \;\mbox{for
RHIC and}\;\approx \fr{1}{19}\;\mbox{ for LHC}\,.
 \label{66a}
  \ee

\begin{figure}[htb]
\includegraphics[width=0.8\linewidth]{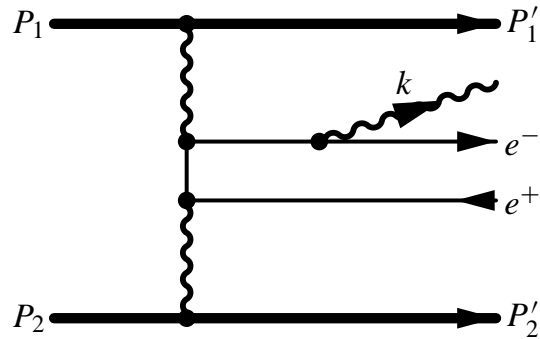}
\caption{Emission of photon by electron or positron,
produced in a nuclear collision.}
 \label{Fig_7}
\end{figure}

%
%
\subsection{Comparison to
$\boldsymbol{ Z_1 \, Z_2\to Z_1 \, Z_2 \, e^+ \, e^- \, \gamma }$}

An alternative source of photons in nuclear collisions
provides by the process
 \be
Z_1 \, Z_2\to Z_1 \, Z_2 \, e^+ \, e^-\, \gamma\,,
 \label{67}
 \ee
in which the photon is emitted by an electron or a positron
produced in collisions of nuclei (Fig.~\ref{Fig_7}). Such a
process has been calculated in the paper \cite{HTB} using
various methods for the numerical evaluation of its cross
section. Strictly speaking, the final state of this process
is different from that of $Z_1 \, Z_2\to Z_1 \, Z_2 \,
\gamma$. Nevertheless, it is of interest to compare the
differential over the photon momentum distribution in this
process with that given by Eq.~(\ref{46}). For this
purpose, we use the analytic expression from
Ref.~\cite{FKh} (derived originally for the $e^+e^-$
collisions), which we denote by $\dd \sigma_{\rm FKh}$, in
order to reflect the names of the authors of the
publication~\cite{FKh},
\begin{align}
\dd \sigma_{\rm FKh} =& \;
\fr{7}{3\pi^3}\,
\fr{\alpha \, (Z\alpha)^4}{k_\perp^4} \,
L^2\,\left(L_0-0.06\right)\,
\fr{\dd^3k}{E_\gamma}\,,
\nonumber\\[2ex]
L_0=& \; \ln\left( \fr{k_\perp^2}{m^2} \right)\,,
 \label{68}
\end{align}
which is valid for $k_\perp^2\gg m^2$.
For Au--Au collisions at the RHIC collider,
the ratio
\begin{align}
& \fr{\dd \sigma_{\mbox{\scriptsize Eq.~(\ref{46}})}}{\dd
\sigma_{\rm FKh}} = \fr{6\pi}{7} \,\fr{(Z\alpha)^2\,
f_Z(k_\perp/m)}{L\,\left(L_0-0.06\right)} \approx
\fr{1}{25} \,,
\nonumber\\[2ex]
& \mbox{for} \qquad k_\perp^2=20\, m^2\,.
\label{69}
\end{align}

In other words, the mechanism of the associative production of
photons as shown in Fig.~\ref{Fig_7} looks much more efficient than
that considered in this paper. However, the estimate
(\ref{69}) is quite rough, since the analytic expression
(\ref{68}) is valid for large photon emission angles, but
our equation (\ref{46}) is correct for small emission
angles. It should also be mentioned that numerical
estimates in the paper~\cite{HTB} give results which are by
one order of magnitude smaller that those given by Eq.
(\ref{68}) (see, for example, Fig. 5 in Ref.~\cite{HTB}).
It means that the yield of photons in the process
Fig.~\ref{Fig_7} exceeds that in the process considered not
so strong. Due to the difference of the final states in these
processes, we can conclude that under a
suitable differentiation of events by detectors,
our virtual Delbr\"{u}ck scattering process can be observed
if a reasonable efficiency of electron or positron recording
can be achieved.

%
%
\section{Delbr\"uck scattering and
the $\boldsymbol{ e\, Z\to e\, Z \, \gamma }$ process}
 \label{scatteringeZ}

In this section we discuss briefly the process of a photon
emission in electron-nucleus collisions without excitation
of nucleus in the final state:
 \be
e(P_1)+Z(P_2) \to e(P^\prime_1)+Z(P^\prime_2) +\gamma(k)\,.
 \label{70}
 \ee
As it was mentioned in Introduction, this consideration is
motivated by project of the eRHIC collider
(see~\cite{eRHIC}) with parameters:
 \be
Z_1=-1\,,\;\;\gamma_1=2\cdot 10^4\,,\;\;Z_2\equiv
Z=79\,,\;\; \gamma_2=108\,.
 \label{71}
 \ee

The process (\ref{70}) via virtual Compton scattering
and via virtual Delbr\"uck scattering is described by
Feynman diagrams Fig.~\ref{Fig_1} and Fig.~\ref{Fig_2},
respectively, in which the first nucleus is replaced by the
electron. The corresponding calculations are basically the
same as above with some minor changes. In particular, in
Eq.~(\ref{14}) we can neglect not only $\sigma_2$, but
$\tilde\sigma_1$ as well:
 \be
\dd\sigma=\dd \sigma_1 =\dd n_e \,\sigma_{\rm D}(Z)
\label{72}
 \ee
with (see, e.\,g., Ref.~\cite{BGMS})
 \bea
\dd n_e&=&2\,{\alpha \over \pi}\,\left(1-x+\fr 12 x^2
\right) \,{\dd\omega \over \omega}\, \ln\left(
\fr{m\gamma_1\sqrt{1-x}}{\omega} \right)\,,
 \nn\\
x&=&\fr{\omega}{E_e}\,.
 \label{73}
 \eea
The total cross section is
 \be
\sigma = \fr{\alpha}{\pi}\, \sigma_{\rm D}(Z)\,L^2\,,
 \label{74}
 \ee
which leads to the value
 \be
\sigma= 0.19\times 10^{-3}\;\;\mbox{ barn}
 \label{75}
 \ee
for the parameters given in Eq.~(\ref{71}).

However, the main interest to this process is connected not
with the total cross section, but with the spectrum of
photons flying along the direction of the nuclear beam,
i.\,e. in the subregion $A$ [see Eq.~(\ref{38})]; this
spectrum is given by
\begin{align}
& \dd \sigma^{(A)}= \fr{2\alpha}{\pi}\, \sigma_{\rm D}(Z)
\ln {\fr{\gamma_1E_\gamma}{m}}\,  \fr{\dd
E_\gamma}{E_\gamma}\,,
\nonumber\\[2ex]
& m\ll E_\gamma\ll m \, \gamma_2 \,.
 \label{76}
\end{align}
By contrast, the spectrum of the ordinary nuclear brems\-strahlung
in the same direction and the same region of energy has the form
 \be
\dd \sigma_{\rm br}^a= \fr{2\alpha}{\pi}\, \, \sigma_{\rm
T}(Z)\, \ln\left( \fr{4\,\gamma_1 \, \gamma_2^2}{E_\gamma
\, R} \right)\; \fr{\dd E_\gamma}{E_\gamma}\,,
 \label{77}
 \ee
and the integrated contribution of this spectrum from the
photon energy region $m< E_\gamma< m \, \gamma_2$ is small,
\be
\fr{\Delta\sigma_{\rm br}^a}{\sigma^{(A)}} \approx \,\fr{1}{15}\,,
\ee
again for the eRHIC parameters given in Eq.~(\ref{71}).

In the subregion $B$ [see Eq.~(\ref{39})], the photons fly
along the electron beam direction, and their spectrum is
\begin{align}
\dd \sigma^{(B)}=& \; \fr{2\alpha}{\pi}
\sigma_{\rm D}(Z)
\left(1-x_\gamma+\fr 12 x_\gamma^2 \right) \,
\nonumber\\[2ex]
& \times \ln \left( {\fr{m\gamma_1\sqrt{1-x_\gamma}}{E_\gamma}} \right)
\fr{\dd E_\gamma}{E_\gamma}\,,
\label{78}
\end{align}
for $ m\ll E_\gamma\ll m \, \gamma_1$, and we use the notation
$x_\gamma =E_\gamma/E_e$. However, in this direction the ordinary
brems\-strahlung is absolutely dominant, since its spectrum, given
by well-known Bethe--Heitler formula (see, for example,
Ref.~\cite{BLP}), is determined by the Compton scattering off the
electron:
 \bea
\dd \sigma_{\rm br}^b&\approx&
\fr{16}{3}\fr{Z^2\alpha^3}{m^2}\,\left(1-x_\gamma+\fr 34\,
x_\gamma^2 \right) \, L_\gamma\, \fr{\dd
E_\gamma}{E_\gamma}\,,
 \nn\\[2ex]
L_\gamma&=&\ln{\fr{4\,\gamma^2_1
\gamma_2\,(1-x_\gamma)}{x_\gamma}} \,.
 \label{79}
 \eea

At the end of this section we consider, for completeness,
the contribution of the virtual Delbr\"uck scattering to
emission of photons in the collision of a high-energy muon
with a heavy nucleus at rest. Let $\mu$ and $\gamma_\mu$ be
the mass and the Lorentz factor of the muon, respectively.
In this case, the emission of photons via virtual
Delbr\"uck scattering is described by the cross section
\begin{align}
& \dd \sigma= \fr{2\alpha}{\pi} \sigma_{\rm D}(Z)\,
\ln \left( {\fr{m\gamma_\mu}{E_\gamma}} \right) \,
\fr{\dd E_\gamma}{E_\gamma}\,,
\nonumber\\[2ex]
& m\ll E_\gamma\ll m \, \gamma_\mu\,,
\end{align}
while the ordinary nuclear bremsstrahlung in the same
region of energy has the form
 \be
\dd \sigma_{\rm br}\approx
\fr{16}{3}\fr{Z^2\alpha^3}{\mu^2}\,
\ln\left( {\fr{\gamma^2_\mu}{RE_\gamma}}\right)\,
\fr{\dd E_\gamma}{E_\gamma}\,.
 \ee
As a  result, for muon-nucleus collisions, the emission of
photons via a virtual Delbr\"uck scattering is small
compared to the ordinary muon bremsstrahlung. E.\,g., even
for a Uranium nucleus, the ratio $\dd \sigma/\dd
\sigma_{\rm br}$ is only about 2\% for $\gamma_\mu>10^3$.

%
%
\section{Conclusions}
\label{conclusions}

We have considered photon emission in collisions of
ultra-relativistic heavy nuclei via the virtual Delbr\"uck
scattering subprocess. Although our analysis has been more
general, for reasons of clarity, we focus on the case of a
symmetric collision with $\gamma = \gamma_1 = \gamma_2$ and
$Z = Z_1 = Z_2$ in this summary. The emitted photon energy
region we consider is
\begin{equation}
{m} \ll E_\gamma \ll m \,\gamma
\end{equation}
in the collider reference system. In the leading
logarithmic approximation, the total photon emission cross
section corresponding to this photon energy region is given
by Eq.~(\ref{27}) as
 \be
\sigma = \fr{2 \alpha}{\pi}\, Z^2 \, \sigma_{\rm D}(Z) \,
\left[\ln(\gamma^2)\right]^2\,,
 \ee
where the quantity $\sigma_{\rm D}(Z)$ is related to the
high-energy limit of the Delbr\"{u}ck scattering cross
section according to Eq.~(\ref{7}). Note that the factor
$r_Z$ in Eq.~(\ref{7}) takes care of the reduction of this
cross section by the Coulomb corrections, which are of
relative orders $\sim (Z\alpha)^{2n}$, and which reduce the
lowest-order result by almost a factor two.
Equation~(\ref{27}), in a certain sense, constitutes the
main result of this paper. Corrections to this result are
of the order of $1/\ln(\gamma^2) \approx 0.06$ for the LHC Pb--Pb
option.

The energy and angular distribution of photons, emitted due
to the Delbr\"uck subprocess, is given by Eq.~(\ref{46}) in
the leading logarithmic approximation, where we recall that
the function $f_Z(k_\perp/m)$ is slowly varying in terms of
$k_\perp/m$, and that numerical values of this function can
be found in Refs.~\cite{Milstein-Schumachar-94,F-92}.

The cross section due to the virtual electron loop as given
in Fig.~\ref{Fig_2} is found to be considerably larger than
that for the usual ``tree-level'' nuclear bremsstrahlung
given in Fig.~\ref{Fig_1}. The ratio of these cross
sections can be found in Eq.~(\ref{66}), and a numerical
result is given in Eq.~(\ref{66a}). Figuratively speaking,
we can say that the considered process depicted in
Fig.~\ref{Fig_2} represents a QED radiative loop correction
which is by one order of magnitude larger then the
tree-level process of Fig.~\ref{Fig_1}.

If our process can be detected experimentally, then one can
effectively study Delbr\"uck scattering in the range of
initial photon energies in the rest frame of the
colliding nucleus up to $\omega_L\sim 2m\gamma_1\gamma_2 $,
which is 10~GeV for RHIC, 8~TeV for LHC and 2~TeV for
eRHIC.

%
%
\section*{Acknowledgments}

We are grateful to G.~Baur, V.~Fadin and A.~Milstein for
useful discussions. V.G.S.~acknowledges the warm
hospitality of the Institute of Theoretical Physics of
Heidelberg University and support by the Gesellschaft
f\"{u}r Schwerionenforschung (GSI Darmstadt). This work is
partially supported by Russian Foundation for Basic
Research (code 06-02-16064) and by Fund of Russian
Scientific Schools (code 5362.2006.2). U.D.J.~acknowledges
support by Deu\-tsche Forschungsgemeinschaft (Heisenberg
program).


\end{document}